\documentclass[AMA,STIX1COL]{WileyNJD-v2}
\usepackage{moreverb}
\usepackage{CJKutf8}

\newcommand\BibTeX{{\rmfamily B\kern-.05em \textsc{i\kern-.025em b}\kern-.08em
T\kern-.1667em\lower.7ex\hbox{E}\kern-.125emX}}

\articletype{Article Type}%

\received{\color{red}Added at production}
\revised{\color{red}Added at production}
\accepted{\color{red}Added at production}

\begin{document}

\title{A machine learning method correlating pulse pressure wave data with pregnancy}

\author[1]{Jianhong Chen}

\author[2]{Huang Huang}

\author[1]{Wenrui Hao}


\author[1]{Jinchao Xu*}

\authormark{Jianhong Chen \textsc{et al}}

\address[1]{\orgdiv{Department of Mathematics}, \orgname{Pennsylvania State University}, \orgaddress{\state{Pennsylvania}, \country{USA}}}

\address[2]{\orgdiv{School of Mathematical Sciences}, \orgname{Peking University}, \orgaddress{\state{Beijing}, \country{China}}}

\corres{Jinchao Xu, Department of Mathematics, Pennsylvania State University, University Park, Pennsylvania, USA. \email{jxx1@psu.edu}}

\begin{CJK*}{UTF8}{gbsn}
  \abstract[Abstract]{ Pulse feeling (中医把脉), representing the
    tactile arterial palpation of the heartbeat, has been widely used
    in traditional Chinese medicine (TCM) to diagnose various
    diseases.  The quantitative relationship between the pulse wave
    and health conditions however has not been investigated in modern
    medicine.  In this paper, we explored the correlation between
    pulse pressure wave (PPW), rather than the pulse key features in
    TCM, and pregnancy by using deep learning technology.  This
    computational approach shows that the accuracy of pregnancy
    detection by the PPW is 84\% with an AUC of 91\%.  Our study is a
    proof of concept of pulse diagnosis and will also motivate further
    sophisticated investigations on pulse waves.}



\maketitle


\section{Introduction}\label{sec1}

\begin{figure}[!htbp]
  \centering
   \includegraphics[width=.65\textwidth]{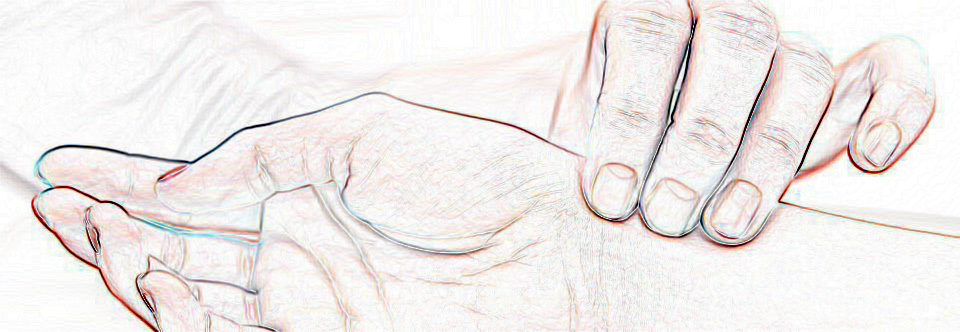}
\includegraphics[width=0.3\textwidth]{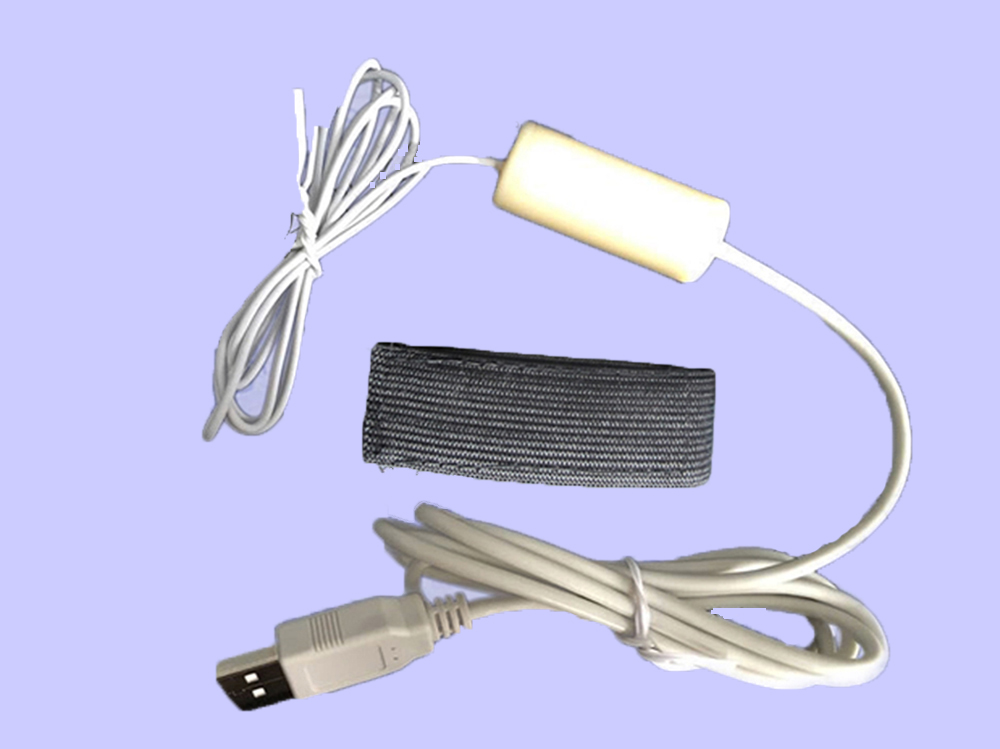}
\caption{Traditional pulse-feeling (left) and pulse pressure sensor (right)}
\label{fig:pulse-feeling}
\end{figure}

Pulse-feeling, obtained by putting the doctor's fingers on a patient's
wrist pulse (see Fig. \ref{fig:pulse-feeling}), has been widely used
in Traditional Chinese Medicine (TCM) for thousands of
years~\cite{shu2007developing,lu2012integrative}.  It has long been
claimed that the pulse feeling can be used to detect various health
conditions such as kidney failure~\cite{arulkumaran2010pulse}, liver
fibrosis~\cite{ghany2005assessment}, cardiovascular
disease~\cite{safar2003current}, and pregnancy~\cite{khalil2009pulse}.
Every TCM doctor is required to master pulse-feeling, which is a very
basic technique, but it is hard to teach and learn.  The teaching and
learning process often takes years, or even a lifetime since the
pulse-feeling varies for different health conditions and patients.
The exact mechanism used in the pulse-feeling is highly complex and
there have been many types of descriptions and theories in Chinese
medical literature, but many of such theories are often be more
subjective than objective.  One such theory is to characterize a
pulse-feeling through three key features ~\cite{tang2012digitalizing}:
length, depth, and pattern.  The length is related to the strength and
balance of the blood and energy flow, the depth reflects certain types
of pathological condition, and the pattern refers to different phases
or features of a pulse-feeling (it is usually believed that there are
28 different pulse patterns \cite{tang2012validation}, such as the
so-called choppy and slippery pulses).

Naturally it is a subject of research and practical interest if the
pulse-feeling technique can be studied and understood through a
scientific manner.  One can imagine this is an extremely challenging
task due to the ``subjective" aspect of the pulse-feeling process.
There are many different approaches in these
studies.
%
%
One such approach is to analyze the aforementioned three
key features of pulse-feeling~\cite{tang2012digitalizing,
  tang2012validation, sun2005multilevel} via various computational methods, e.g.,  artificial neural network (ANN) developed in recent years.
By using certain physical parameters of arterial pressure waveform
acquired from six locations (left and right {\em cun, guan}, and {\em
   chi}) as inputs, an ANN with one
 hidden layer of width $45$ has been employed \cite{tang2012digitalizing}
for a regression study of features of pulse-feeling. The number of
 data used in the study was $n=229$ and the value of $R^2$, a standard
 parameter measuring model performance, for their ANN model
was $0.6-0.86$. Another similar ANN with a hidden layer of width $25$
has also been developed \cite{tang2012validation}  to classify normotension versus
hypertension. This study was based on certain pulse assessment, namely, the intensity of features of pulse-feeling measured by TCM experts, on a dataset with
 $139$ normotension and $121$ hypertension.  Although the accuracy was $80\%$,  the
conclusion in this study was more subjective than objective because features of pulse-feeling highly rely on the doctor's experience. Thus, the conclusions reached by this approach  may vary among different doctors and may even vary
when given by the same doctor who examines the same patient but at a
different time or in a different environment.

\begin{figure}[!htbp]
  \centering
  \includegraphics[width=.8\textwidth]{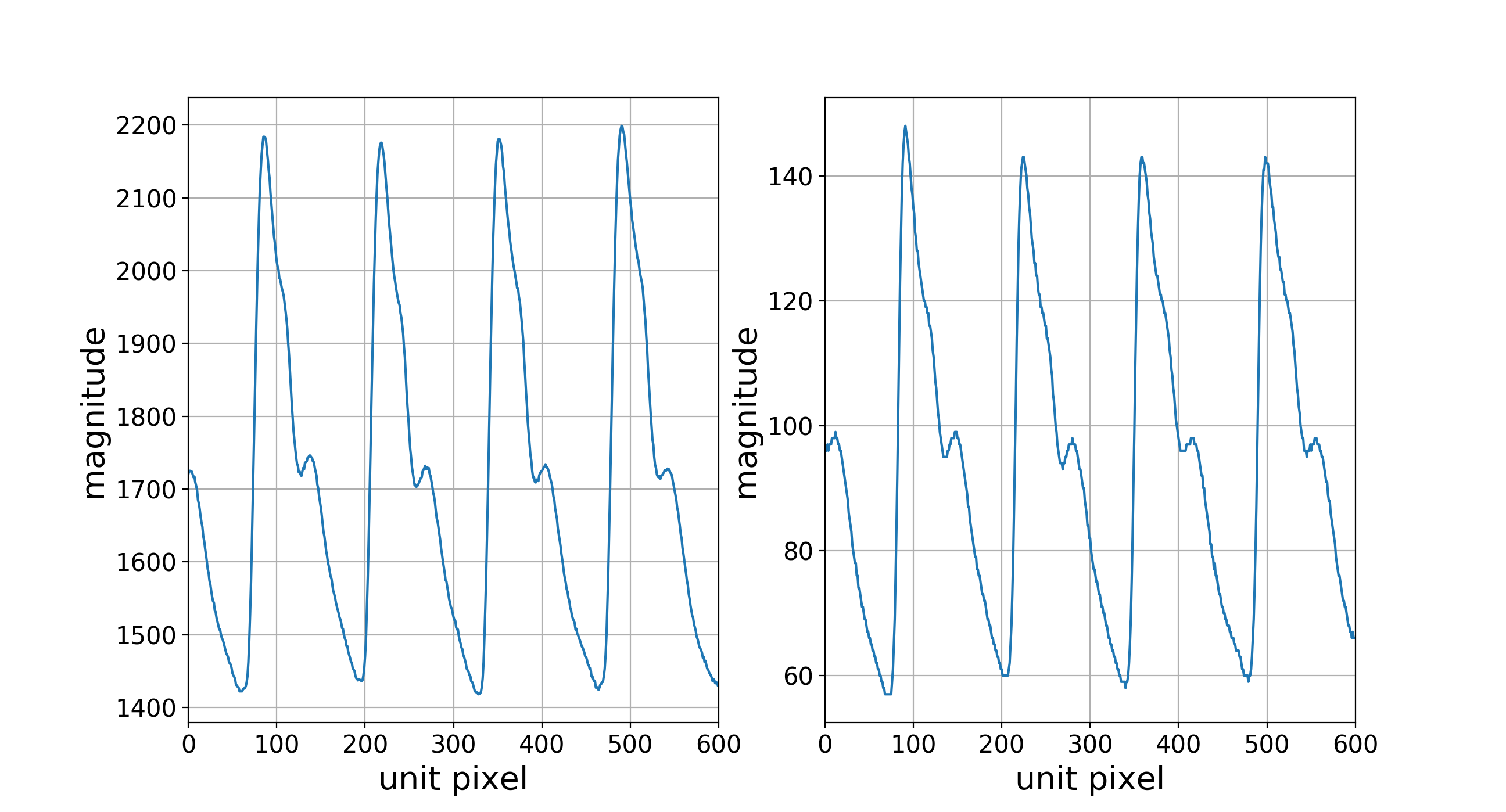}
  \caption{{\bf Sample data:} the PPW data measured during 4 seconds (left); the PPG data measured during the same periods (right). }
  \label{sample}
\end{figure}

Another approach is to use an appropriate medical device to collect
pulse signals and then quantitatively
interpret these signals for the health diagnosis.
One such  device that has been used by researchers is the pulse pressure
sensor as shown in Fig.~\ref{fig:pulse-feeling} (right).   By placing this
type of sensor on a patient's wrist pulse, pulse pressure waves (PPW),  as
shown in Fig.~\ref{sample} (left), can be collected.
A PPW reflects pressure changes in the wrist blood vessel and is believed
to contain certain information that a TCM doctor uses in
pulse-feeling diagnosis. Then a quantitative study can be carried
out by analyzing these PPWs.
%
%
For example, %
by using PPW as the input, a 12-pulse-pattern classification (such as
stringy, slippery, etc) was conducted with a 9-layer 1D {convolutional
  neural network (CNN)} (with around 2,280 variables). On 200
training and 261 test samples, they obtained a 93.49\%
accuracy~\cite{zhang2016human}.
Moreover, a 9-layer 1D CNN was used to further study the correlation
between PPW and arteriosclerosis.  60\% of the data was used as
training data and 40\% was used as test data from a dataset with 47
participants (35 arteriosclerosis and 12 non-arteriosclerosis) and
obtained a 96.33\% accuracy rate on classifying the arteriosclerosis
versus {non-arteriosclerosis~\cite{hu2014wrist}.}
The relationship between important features of the arterial pulse wave
(peak, length, etc) and pregnancy has also been studied through using a
4-hidden layer probabilistic neural network (the nodes for each layer
are 22, 44, 2, and 1) ~\cite{wang2008recognition}. By using 110
samples of training data (45 pregnant and 65 non-pregnant), the neural
network reached 100\% accuracy on a test {dataset} with 20 samples (4
pregnant and 16 non-pregnant).
It is interesting to note that all these studies were based on small
datasets (around 200 samples), although the number of variables was
relatively large (around 2,000). Moreover, the training/test datasets
were imbalanced (i.e., pregnant/non-pregnant ratio is 2/3 in the
training data but is 1/4 in the test data \cite{wang2008recognition}),
and there was a lack of cross-validation, which is important for a
generalizable computational model.


Other techniques related to PPW are also commonly used in modern
medical practices.  The most directly related technique is the pulse
wave velocity (PWV) that has been used as an index of arterial
stiffness~\cite{sutton2005elevated} and a biomarker of cardiovascular
risk~\cite{blacher1999aortic} and even other
diseases~\cite{blacher2003aortic}. Another relevant technique is based
on photoplethysmogram (PPG), which is an optically obtained
plethysmogram that can be used to detect blood volume changes in a
microvascular bed of tissue.  A PPG signal, as shown in
Fig.~\ref{sample} (right), is similar to a PPW signal as shown in the
left panel of Fig~\ref{sample}.
PPG has been widely used in modern clinical practice as a non-invasive
diagnostic technique~\cite{yoon2002multiple,ko2000method} and also for monitoring heart
rate~\cite{zhang2015photoplethysmography}, cardiac cycle,
respiration~\cite{daimiwal2014respiratory} and other bodily processes.

In this paper, we will use deep learning techniques, more specifically CNN,
to study PPW.
There are several points in our study that significantly differ from
the existing studies.  First, our study is based on a relatively large
clinical dataset that was collected from over 4,000 women  consisting
of over 1,800 pregnant and 2,200 non-pregnant women.  In
comparison, existing studies were based on much smaller datasets (for
example, 6 samples
\cite{shi2017evaluation}, 47 samples \cite{hu2014wrist},  and 130 samples \cite{wang2008recognition}).
%
Second, in this paper, we developed a modern computational approach
based on CNN with two different neural network models and data
preprocessing methods to analyze the underlying correlation between
the PPW and health conditions. Due to the similarity among PPW, PPG,
and electrocardiogram (ECG), our approach can be generalized to the
diagnosis of human health conditions by other available clinical
imaging data (PPG, ECG, etc) ~\cite{bagha2011real,das2007fragmented}.
Third, we studied the correlation between the PPW and pregnancy as a
proof of concept due to the following reasons: 1) the status of
pregnancy is clear and definite, unlike kidney weakness
\cite{gong2015th17} and neurasthenia \cite{sandiford1920effect}, which
are vague and unclear; 2) the label of pregnancy is very accurate due
to standard and effective techniques for pregnancy detection, such as
urine tests (with high accuracy) and blood
tests~\cite{wide1960immunological}; 3) it has been widely claimed in
TCM that pregnancy can be accurately detected by pulse
feeling~\cite{liao2012pulse}; 4) during pregnancy, there are many
physiological changes in the reproductive system, circulatory system,
urinary system, respiratory system, etc. Therefore the maternal blood
volume increases by 45\% above non-pregnant
values~\cite{soma2016physiological}, which may change pulse wave
patterns. By taking the original PPW segments as the only input, our
models provided an 85\% accuracy of pregnancy prediction.  The paper is
organized as follows: first, we introduce the data used in our study
and two data preprocessing approaches; second, we present our
computational approach, which consists two different neural networks;
the detailed results for both approaches and discussion are then
presented..

\section{The available dataset \& preprocessing}
\label{sec:data}
The PPW data was collected from over 4,000 people (age $28.69\pm
8.60$) in Nanjing Huashijiabao Hospital, China.  The collections were
taken under the HK-2010/1 single channel pulse sensor, by experienced
and trained nurses, from volunteers' left wrists when the pulse
repeated in a regular pattern for about 40 seconds. {The pulse sensor
  measured the blood pressure with a time resolution of $1/150$
  seconds. In Fig. \ref{sample} we have plotted an image with the
  $x$-axis assigned to the measurements while the $y$-axis is assigned
  to the pulse wave amplitude.}  Then our PPW imaging dataset was
selected based on similar periods and amplitudes. The number of people
with respect to all pregnancy weeks is shown in
Fig~\ref{weekcount}. There are two peaks centering around the 12th and
23rd weeks since two major pregnancy examinations happen around these
weeks.
The low quality refers to the irregular pattern and large noise introduced by the external environment and the collecting operation. After this filtering process, only 1,840 samples (of which 645 are pregnant) are left.
 In order to obtain reliable and reproducible results, we designed a data preprocessing approach: the restrict the PPW to one period {and append zeros  to construct the measurement interval. In our study, we chose the measurement interval as $256$ unit pixels since all the periods are smaller than $256$ unit pixels.} By denoting our dataset $\mathcal D = \{\mathbf a_1,\cdots,\mathbf a_d,\cdots\}$ where $\mathbf a_i\in \mathcal{D}$ is one datapoint, we describe the data preprocessing approach to preprocess each data $\mathbf a_i\in \mathcal{D}$ as follows:

\begin{figure}[!htbp]
\centering
\includegraphics[width=.6\textwidth]{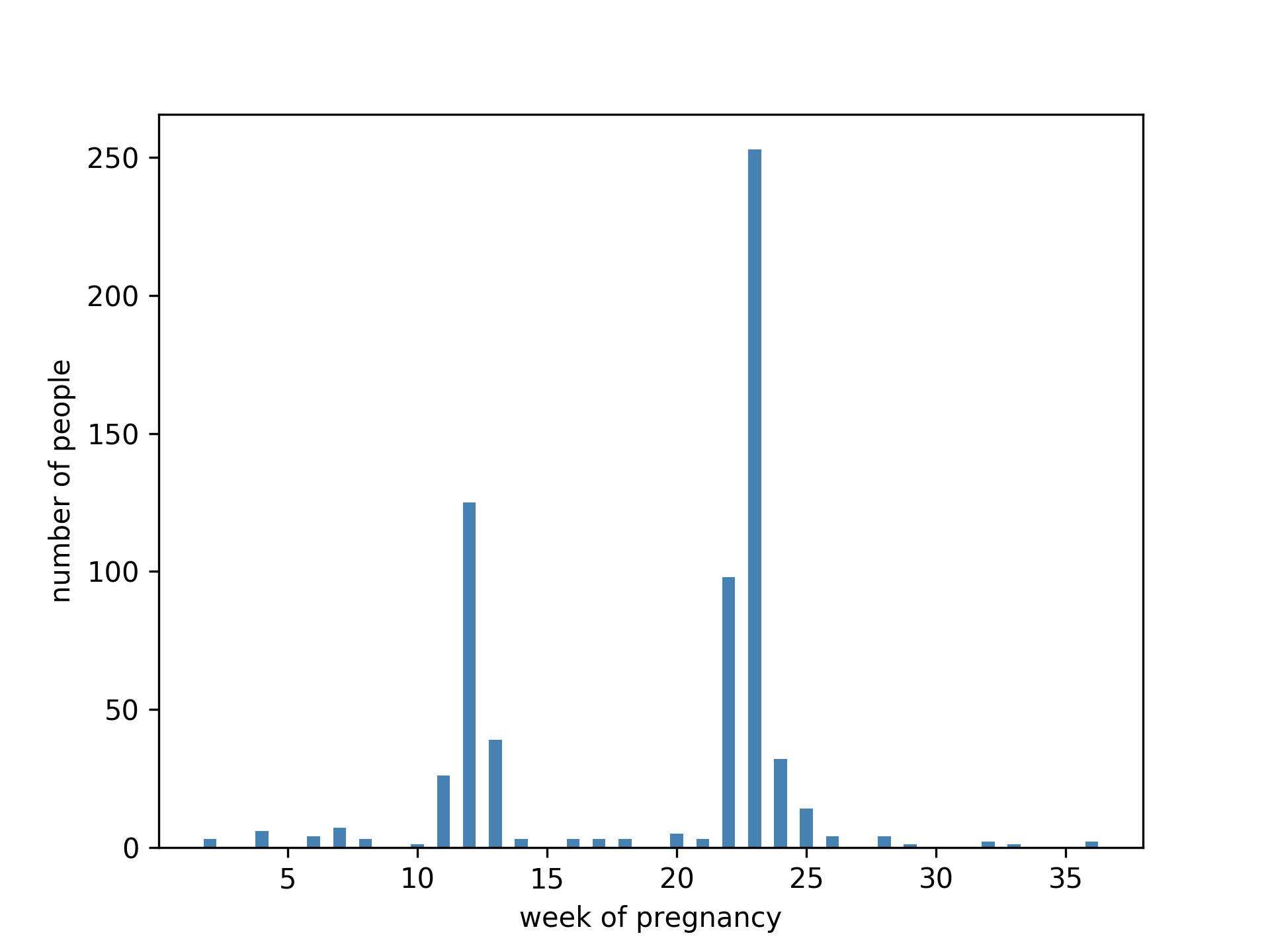}
\caption{{\bf Pregnant week distribution:} the number of female volunteers  in the PPW dataset by pregnancy week.}
\label{weekcount}
\end{figure}

\begin{figure}[!ht]
\centering
\includegraphics[width=.4\textwidth]{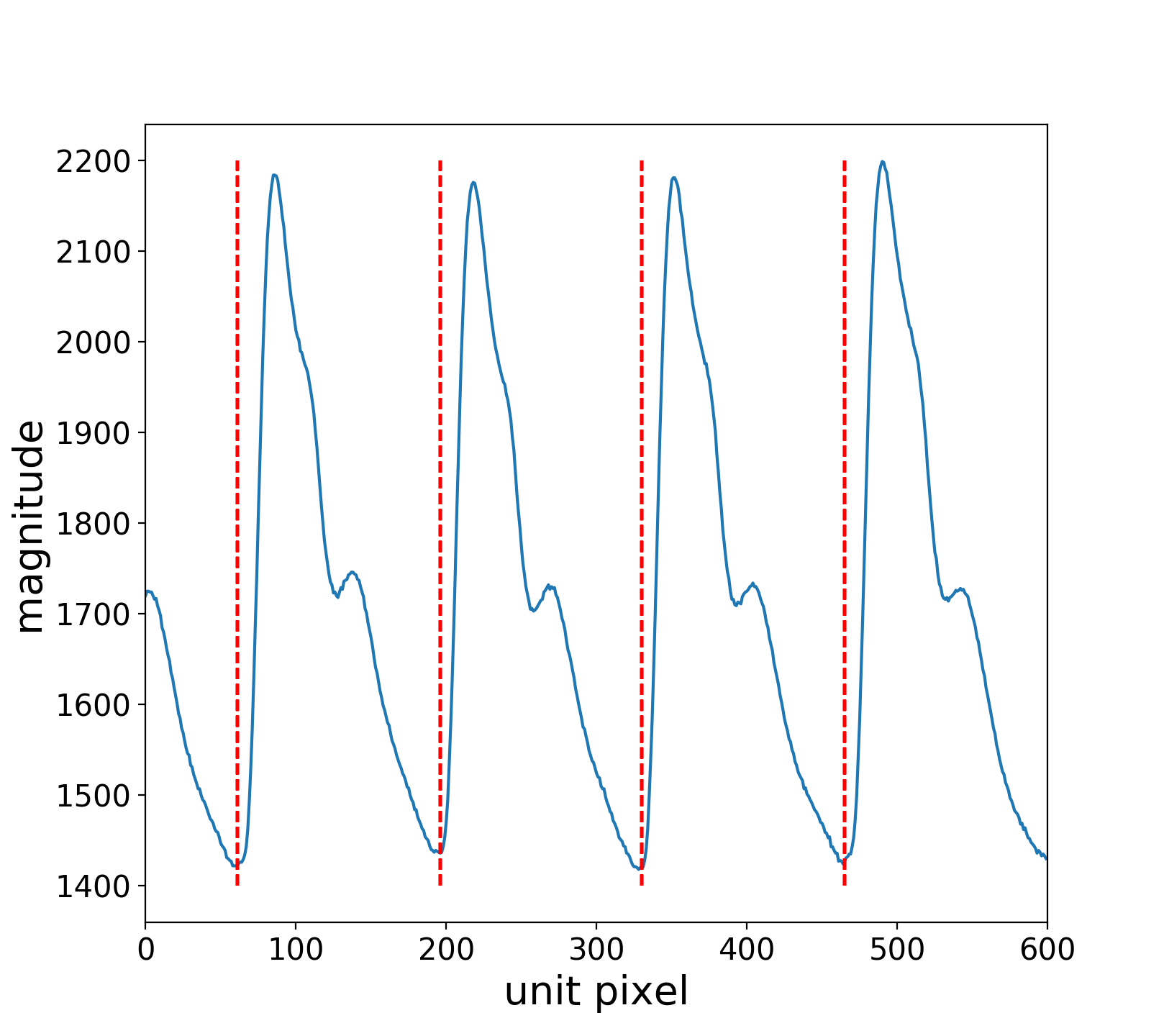}
\includegraphics[width=.4\textwidth]{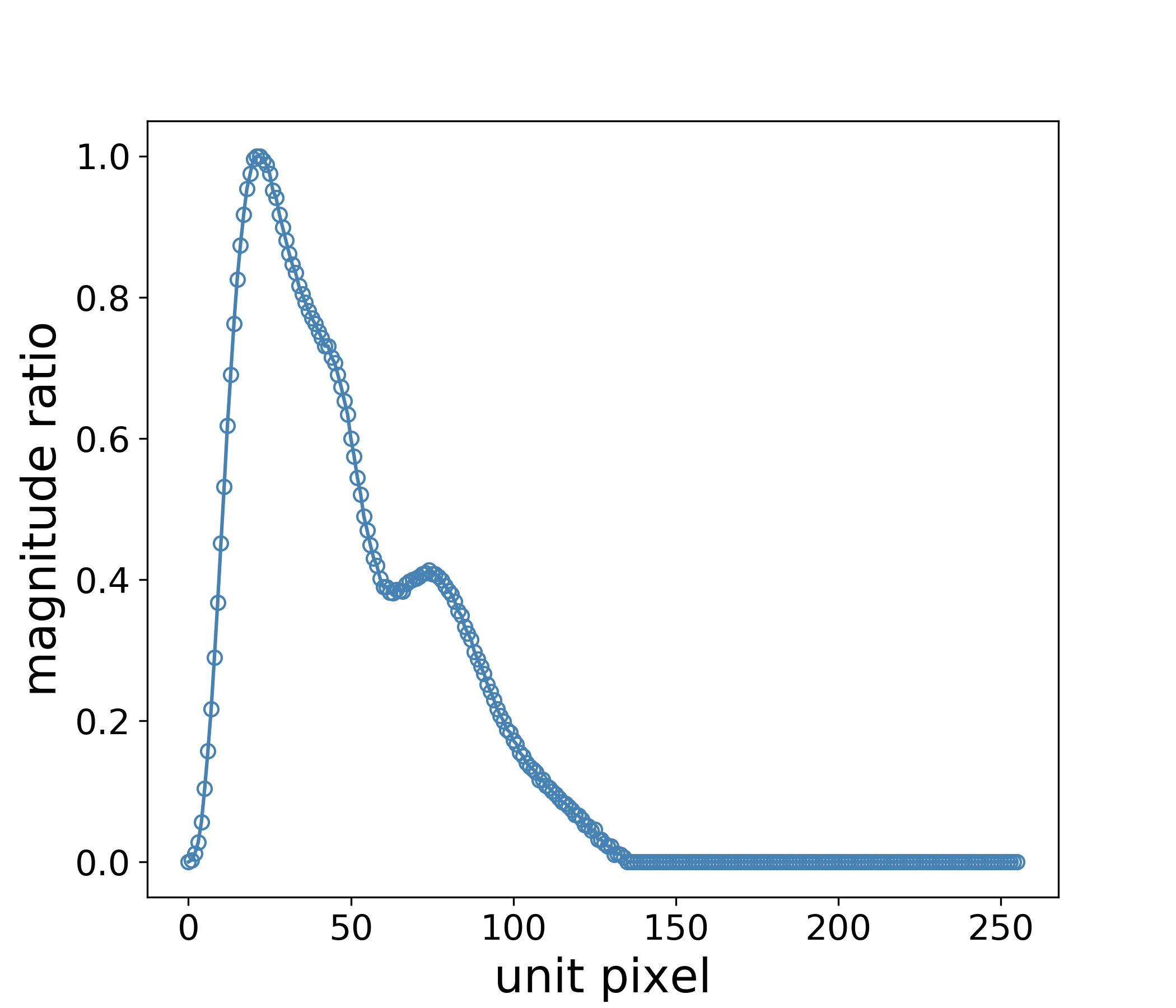}
\caption{The illustration of one period restriction: locating local minimum for sample data (left); one sample period after preprocessing (right).}
\label{preprocess}
\end{figure}

\begin{enumerate}
\item divide $\mathbf a_i$ into several sub-intervals by local minimums, which are shown as red lines in the left part of Fig~\ref{preprocess};
\item and pick one of the sub-intervals in the first step and normalize linearly into [0,1] with 256 pixels (append zeros if less than 256 {since all the periods are less than 256}) shown in the right part of Fig~\ref{preprocess}.

\end{enumerate}

\section{Computational models}
\label{sec:model}
We applied two different types of 1D-CNN models from \cite{he2016deep}
and \cite{juncai2019mgnet} respectively to analyze the PPW data and to
compare the results. Model 1, which is based on the 1D ResNet and uses
BasicBlocks~\cite{he2016deep}, with more channels and parameters (the
number of parameters is 242,642), fewer layers, but a better
generalization capability. The detailed structure of Model 1 is shown
in Table \ref{resnet2}.  Here BasicBlocks, which insert identity
shortcut connections to the plain convolutional layers
\cite{he2016deep}, are commonly used in deep neural networks of the
ResNet to reduce the computational cost.  Model 2 is based on the 1D
MgNet~\cite{juncai2019mgnet} and uses 18 layers with only 8 channels
(the number of parameters is 3,450).  We note that the structures
(and the number of weights used herein) between Model 1 (which is based on the
standard ResNet) and Model 2 (which is based on the new MgNet) are
quite different, but, as we shall see later, the outcomes of these two
different models are similar, which provides a
cross-validation of the liability of using the deep learning approach on the PPW data.


\begin{table}[!htbp]
\caption{The detailed structure of Model 1.}
\label{resnet2}
\centering
		\begin{tabular}{cccc}
			Layer Name& Input Size& Output Size& Layer Parameters\\\hline
			Convolution&$256\times 1 $&$128\times16$& $[7,16]+[16]$ \\
			Pooling&$128\times16 $&$64\times16$& $3$ for max pooling\\\hline
			BasicBlock1& $64\times16$& $64\times16$ & $[3,16]\times 2$\\\hline
			Block with downsample1& $64\times16$& $32\times32$ & $[3,32]\times 3$\\
			BasicBlock2 &$32\times32$&$32\times32$ &$[3,32]\times 2$\\\hline
			Block with downsample2& $32\times32$& $16\times64$ & $[3,64]\times 3$\\
			BasicBlock3 &$16\times64$&$16\times64$ &$[3,64]\times 2$\\\hline
			Block with downsample3& $16\times64$& $8\times128$ & $[3,128]\times 3$\\
			BasicBlock4 &$8\times128$&$8\times128$ &$[3,128]\times 2$\\\hline
			Pooling&$8\times128$&$256$& $7$ for average  pooling\\
			Fully Connected&$256$&$2$& $256\times 2$
	\end{tabular}
\end{table}

\section{Results}\label{sec:result}
\begin{figure}[htbp]
\centering
\includegraphics[width=0.48\textwidth]{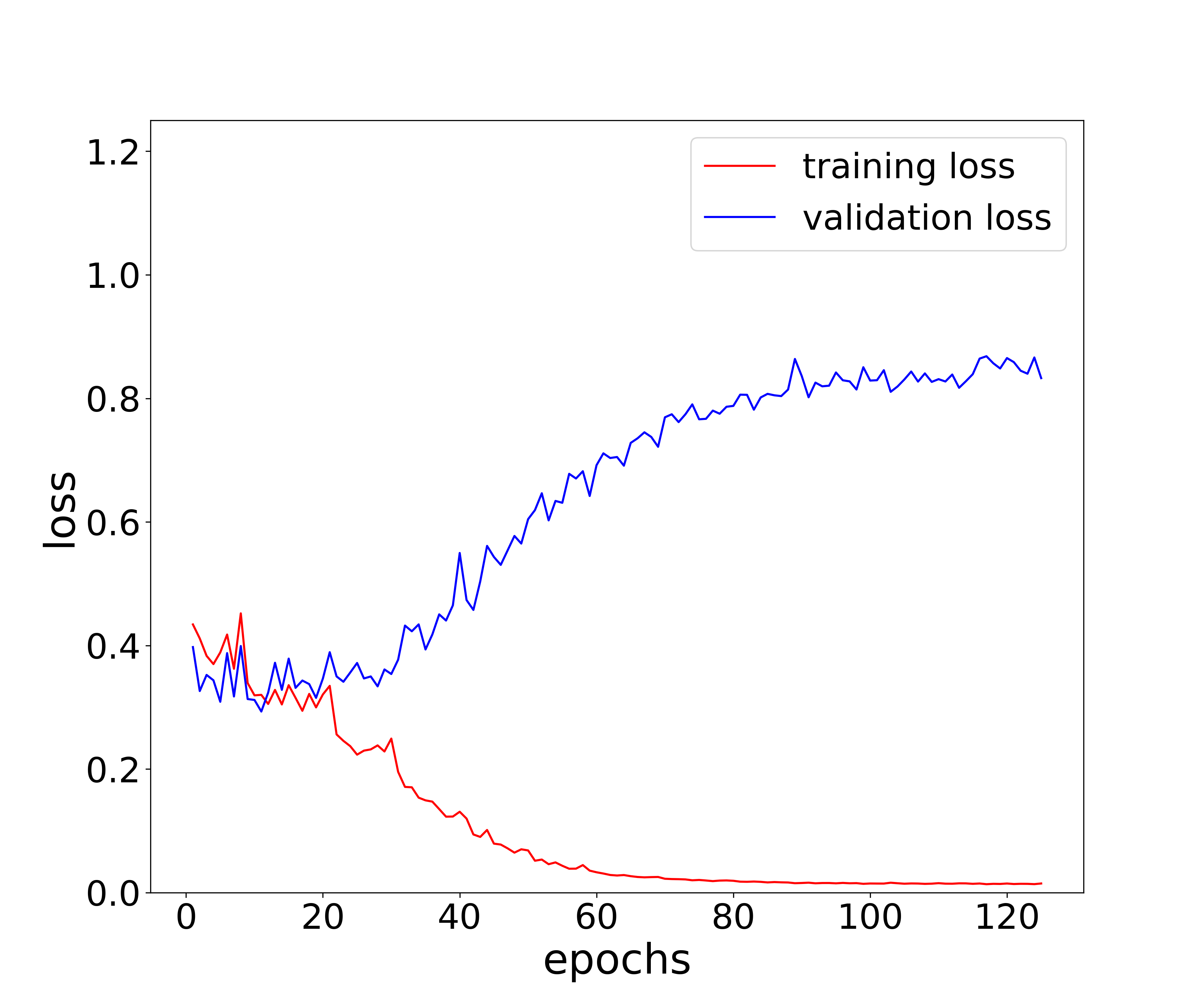}
\includegraphics[width=0.48\textwidth]{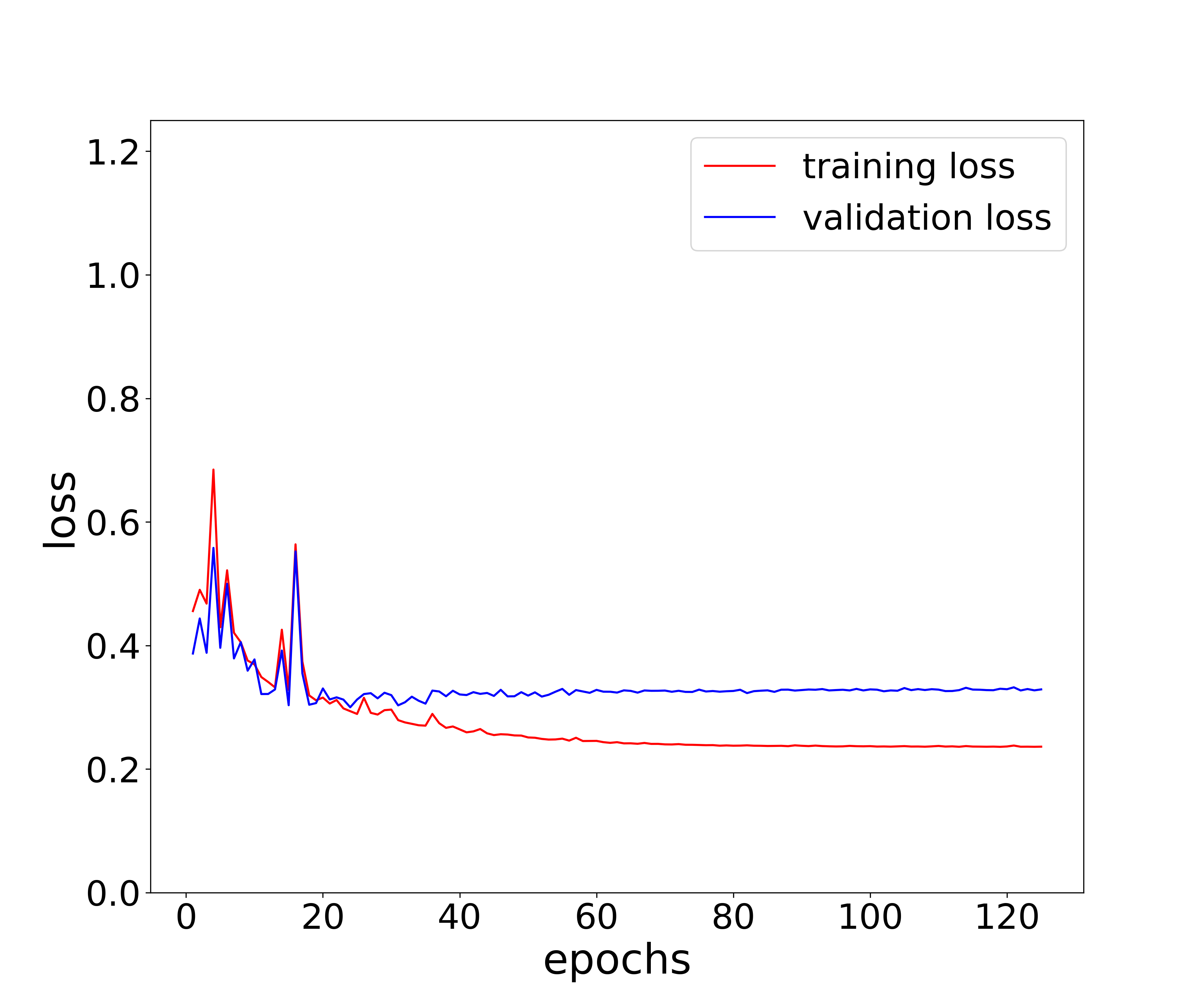}\\
\includegraphics[width=0.48\textwidth]{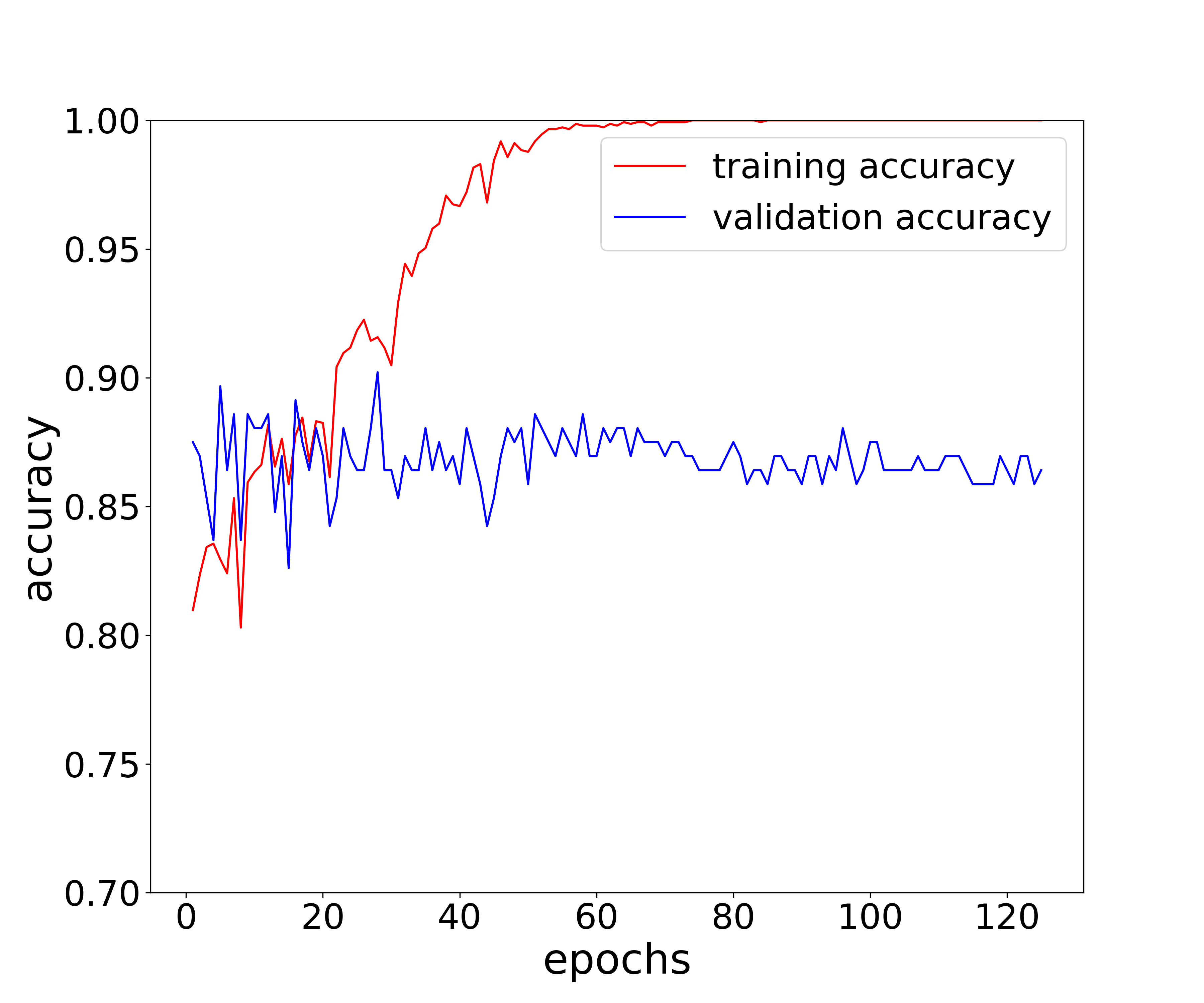}
\includegraphics[width=0.48\textwidth]{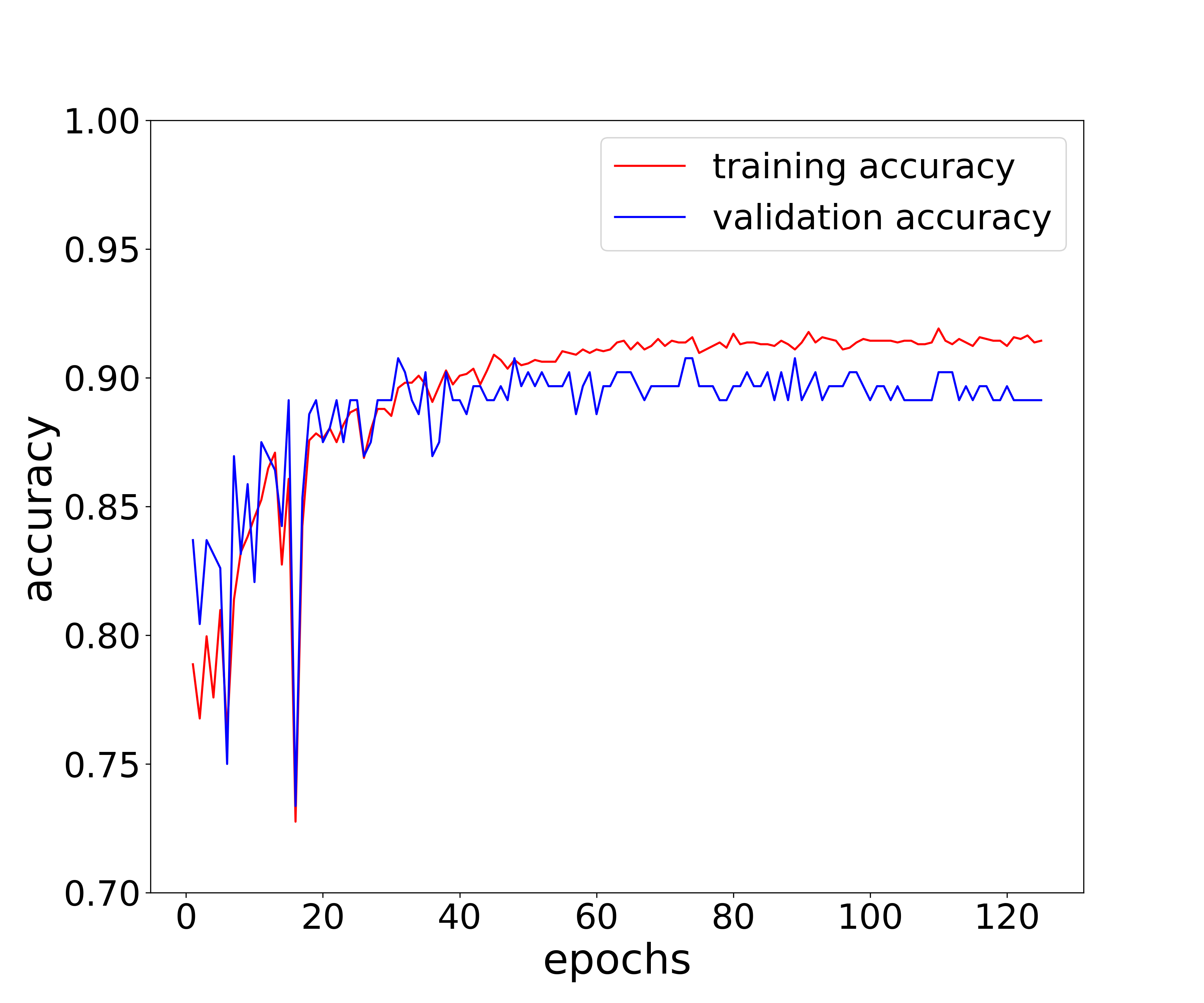}\\
\caption{{\bf Loss and accuracy curves} Left: training/validation loss/accuracy for model 1. Right: training/validation loss/accuracy for model 2}
\label{loss}
\end{figure}

\begin{figure}[htbp]
\centering
\begin{minipage}[t]{0.48\textwidth}
	\includegraphics[width=\textwidth]{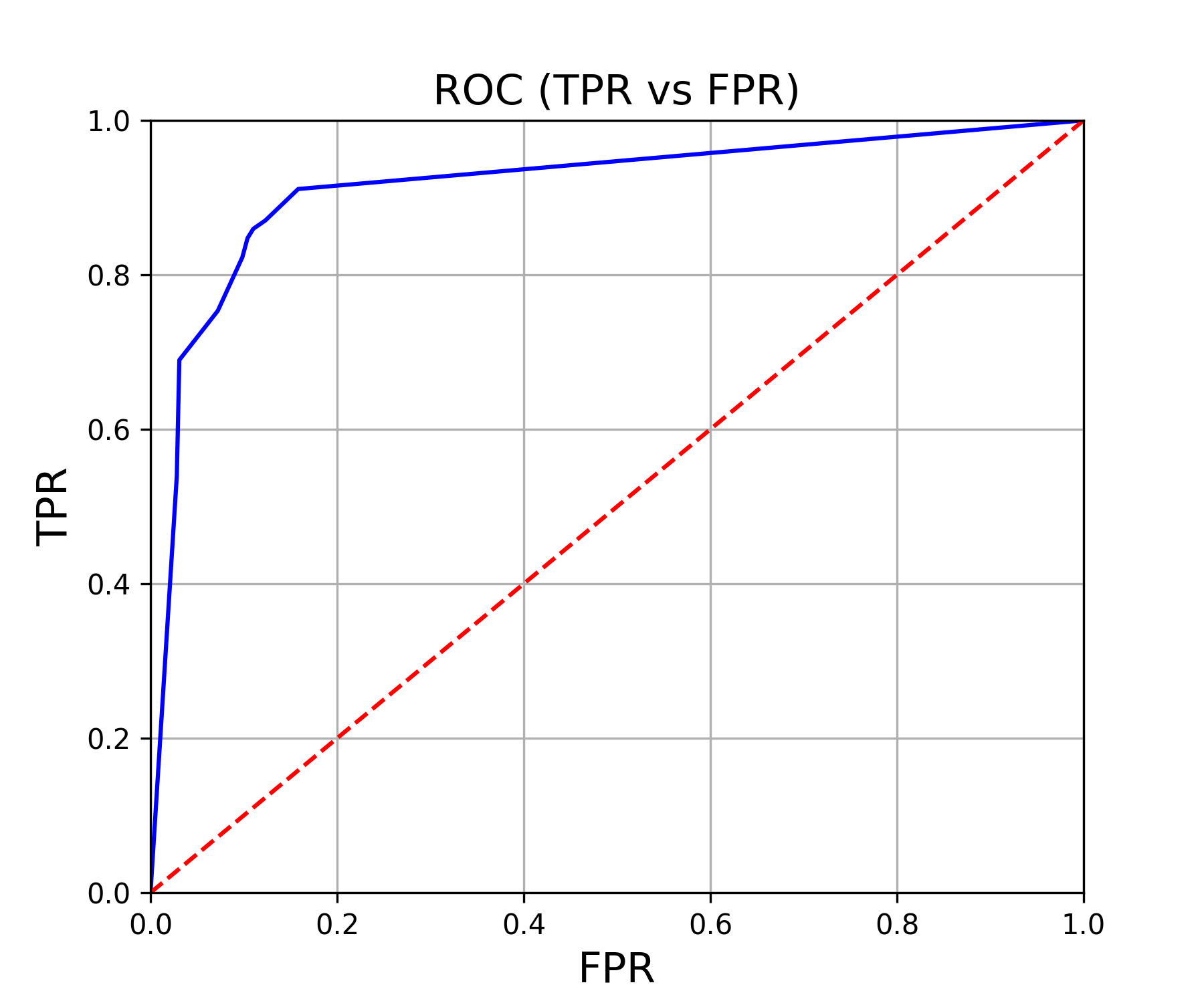}
\end{minipage}
\begin{minipage}[t]{0.48 \textwidth}
	\includegraphics[width=\textwidth]{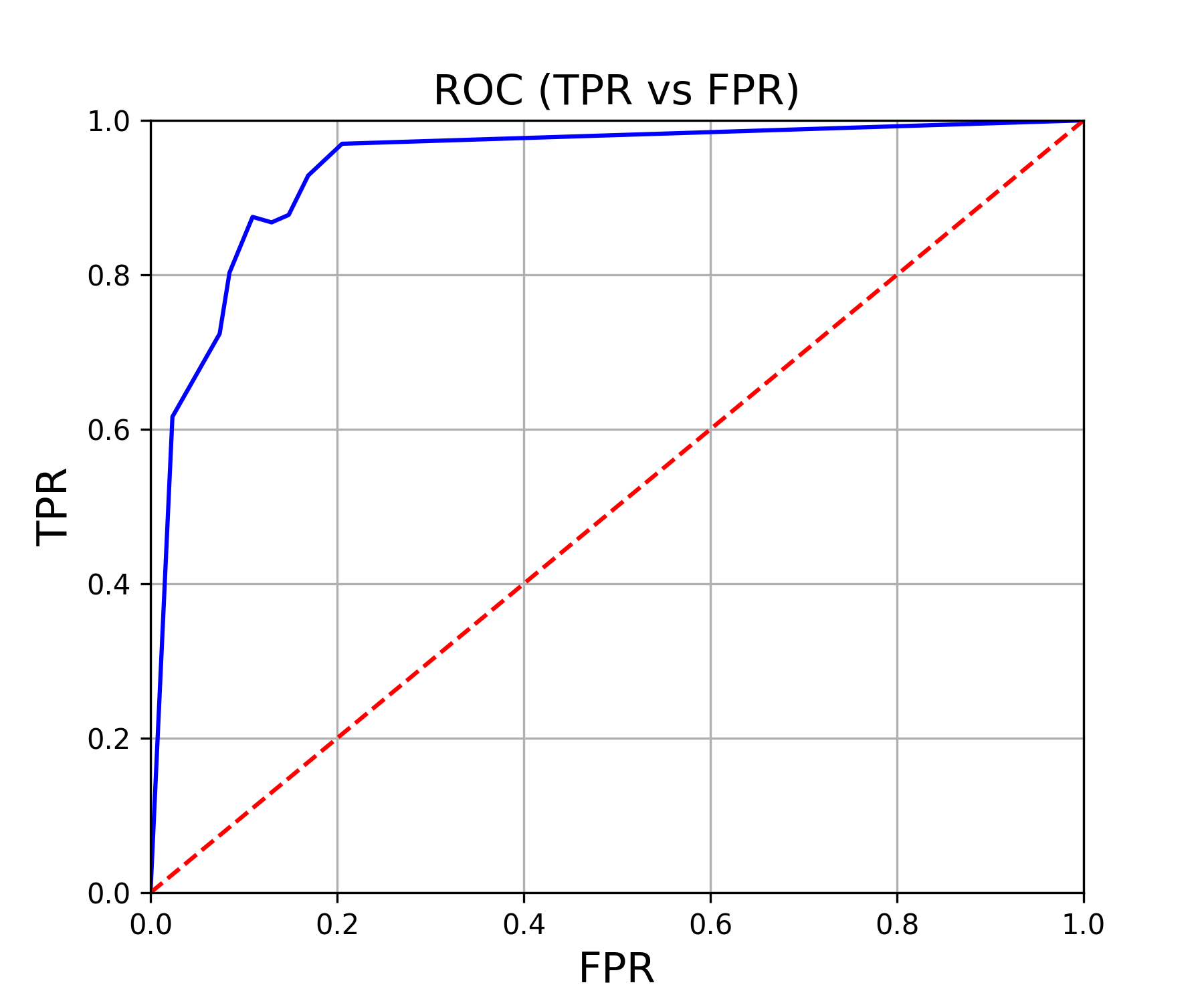}
\end{minipage}
\caption{{\bf ROC curve:} Left: ROC curve  for Model 1. Right: ROC curve for Model 2}
\label{ROC}
\end{figure}

 The implementation of these two computational models is based on
the PyTorch library ~\cite{paszke2017automatic}. The optimization problem is based on the cross-entropy loss, which is defined as
\begin{align}
L(y,\hat{y})=-\frac{1}{n}\sum_i y_i\ln \hat{y}_i +(1-y_i)\ln (1-\hat{y}_i),
\end{align}
where $n$ is the total number of data samples,  $y_i$ is the true label for the $i$-th PPW data, and $\hat{y}_i$ is the  predicted label of the computational models. The Adam method~\cite{kingma2014adam} was employed to solve the optimization problem. {The learning process was stopped after running 200 epochs and then we chose the model with the best validation accuracy.}  The loss and accuracy of the first fold for both models are shown in Fig. \ref{loss}. For Model 1, the average testing accuracy is 84.73\%, while Model 2's average testing accuracy is 84.68\%.
In order to test the generalizability of these computational models, we  used the 10-fold cross-validation to test the classification accuracy of our models. In particular, we divided the dataset (n=1,840) into 10 folds of equal sizes, namely 184 data samples in each fold. The test set was chosen from folds 1 to 10, the validation set was chosen from the next fold, and the remaining data was used as the training set. The cross validation results are shown in Table~\ref{table:cv2} (the left side of each column is for Model 1 while the right side is for Model 2). Moreover, the {receiver operating characteristic (ROC) curve} of the true positive rate (TPR) versus the false positive rate (FPR), an important indicator in statistics, is plotted in Fig~\ref{ROC} {to illustrate the first fold. For all the folds, we computed the test area under the curve (AUC) shown in Table~\ref{table:cv2}.} The average AUC is 89.66\% (Model 1) and 91.04\% (Model 2). 
Finally, we also compared the results of our models with the SVM, the logistic regression method from the Scikit-learn library~\cite{scikit-learn}, and other CNN models~\cite{hu2014wrist,zhang2016human,wang2008recognition}. The accuracy and the AUC for different models are shown in Table~\ref{table:SVM}, which clearly shows that the results of our computational models are more accurate than the others.

\begin{table}[!htbp]
	\centering
	\caption{Accuracy and AUC for two models (Red color for Model 1, while blue color for Model 2.}
	\label{table:cv2}
\begin{tabular}{|c|c|c|c|c|c|c|c|c|c|c|c|}
		\hline
		Test &  \multicolumn{2}{{|c|}}{Training}  &    \multicolumn{6}{{|c|}}{Accuracy}   &\multicolumn{2}{{|c|}}{}  \\\cline{4-9}
		fold &   \multicolumn{2}{{|c|}}{epochs} &    \multicolumn{2}{{|c|}}{Training} & \multicolumn{2}{{|c|}}{Validation}   & \multicolumn{2}{{|c|}}{Testing}    & \multicolumn{2}{{|c|}}{AUC(\%)}\\\hline
1	&	{\color{red}	28	}	&	{\color{blue}	31	}	&	{\color{red}	91.58	}	&	{\color{blue}	89.61	}	&	{\color{red}	90.22	}	&	{\color{blue}	90.76	}	&	{\color{red}	88.04	}	&	{\color{blue}	86.96	}	&	{\color{red}	94.20	}	&	{\color{blue}	95.12	}	\\\hline
2	&	{\color{red}	84	}	&	{\color{blue}	38	}	&	{\color{red}	100.00	}	&	{\color{blue}	90.29	}	&	{\color{red}	84.78	}	&	{\color{blue}	84.78	}	&	{\color{red}	87.50	}	&	{\color{blue}	87.50	}	&	{\color{red}	89.55	}	&	{\color{blue}	91.67	}	\\\hline
3	&	{\color{red}	81	}	&	{\color{blue}	17	}	&	{\color{red}	100.00	}	&	{\color{blue}	89.67	}	&	{\color{red}	85.87	}	&	{\color{blue}	84.24	}	&	{\color{red}	86.41	}	&	{\color{blue}	82.07	}	&	{\color{red}	88.54	}	&	{\color{blue}	89.36	}	\\\hline
4	&	{\color{red}	5	}	&	{\color{blue}	29	}	&	{\color{red}	86.48	}	&	{\color{blue}	90.08	}	&	{\color{red}	87.50	}	&	{\color{blue}	88.59	}	&	{\color{red}	79.35	}	&	{\color{blue}	77.17	}	&	{\color{red}	84.39	}	&	{\color{blue}	85.69	}	\\\hline
5	&	{\color{red}	12	}	&	{\color{blue}	31	}	&	{\color{red}	88.45	}	&	{\color{blue}	90.96	}	&	{\color{red}	88.04	}	&	{\color{blue}	86.96	}	&	{\color{red}	83.70	}	&	{\color{blue}	82.07	}	&	{\color{red}	92.85	}	&	{\color{blue}	91.83	}	\\\hline
6	&	{\color{red}	18	}	&	{\color{blue}	55	}	&	{\color{red}	88.11	}	&	{\color{blue}	89.88	}	&	{\color{red}	91.30	}	&	{\color{blue}	89.67	}	&	{\color{red}	83.70	}	&	{\color{blue}	84.78	}	&	{\color{red}	91.14	}	&	{\color{blue}	89.59	}	\\\hline
7	&	{\color{red}	55	}	&	{\color{blue}	29	}	&	{\color{red}	99.46	}	&	{\color{blue}	92.53	}	&	{\color{red}	86.41	}	&	{\color{blue}	90.22	}	&	{\color{red}	86.96	}	&	{\color{blue}	92.93	}	&	{\color{red}	89.04	}	&	{\color{blue}	93.90	}	\\\hline
8	&	{\color{red}	19	}	&	{\color{blue}	24	}	&	{\color{red}	91.10	}	&	{\color{blue}	90.15	}	&	{\color{red}	85.87	}	&	{\color{blue}	86.96	}	&	{\color{red}	84.24	}	&	{\color{blue}	86.96	}	&	{\color{red}	93.14	}	&	{\color{blue}	91.93	}	\\\hline
9	&	{\color{red}	67	}	&	{\color{blue}	14	}	&	{\color{red}	100.00	}	&	{\color{blue}	86.28	}	&	{\color{red}	89.67	}	&	{\color{blue}	85.87	}	&	{\color{red}	80.98	}	&	{\color{blue}	82.07	}	&	{\color{red}	81.56	}	&	{\color{blue}	92.25	}	\\\hline
10	&	{\color{red}	31	}	&	{\color{blue}	51	}	&	{\color{red}	91.85	}	&	{\color{blue}	89.33	}	&	{\color{red}	89.67	}	&	{\color{blue}	88.59	}	&	{\color{red}	86.41	}	&	{\color{blue}	84.24	}	&	{\color{red}	92.18	}	&	{\color{blue}	89.10	}	\\\hline
avg	&	{\color{red}	40	}	&	{\color{blue}	32	}	&	{\color{red}	93.70	}	&	{\color{blue}	89.88	}	&	{\color{red}	87.93	}	&	{\color{blue}	87.66	}	&	{\color{red}	84.73	}	&	{\color{blue}	84.68	}	&	{\color{red}	89.66	}	&	{\color{blue}	91.04	}	\\\hline
	\end{tabular}
\end{table}

\begin{table}[!htbp]
	\caption{Comparisons of our models with the SVM and logistic regression. }
	\label{table:SVM}
	\centering
	\begin{tabular}{|c|c|c|c|}
		\hline
		Model &  Description & Test accuracy(\%)  & AUC(\%)\\\hline
        1    &    Model 1    &    84.73    &    89.66    \\\hline
        2    &    Model 2    &    84.68    &    91.04    \\\hline
      {\color{red}  3 }   &   {\color{red} CNN 1 \cite{hu2014wrist} }   &   {\color{red} 80.71  }  &  {\color{red}  87.43  }  \\\hline
      {\color{red}  4  }  &  {\color{red}  CNN 2 \cite{zhang2016human}  }     &  {\color{red}  80.76 }   &  {\color{red} 88.19 }   \\\hline
      {\color{red} 5  }  &  {\color{red}  ANN  \cite{wang2008recognition}  }     &  {\color{red}  81.09 }   &  {\color{red} 88.63 }   \\\hline
        6    &    SVM     &    75.54    &    75.72    \\\hline
        7    &    Logistic regression     &    78.80    &    78.24    \\\hline
	\end{tabular}
\end{table}

\section{Conclusion \& discussion}\label{sec:discussion}
In this paper, we have employed
 modern machine learning techniques to explore the relationship between pulse and pregnancy, which has been claimed to exist in TCM over a long history.  Two computational models, based on the 1D CNN, have been developed and have shown that the best accuracy of pregnancy detection by PPW  is 84\% and the best AUC is 91\%.
The computational models we developed  in this paper provide a rigorous and scientific approach to analyze the PPW data.
This approach, based on deep learning, provides a systematic way to ``learn" the correlation between the PPW data and pregnancy.
Although this ``learning" approach, similar to the Alpha-go~\cite{silver2018general},  may not be able to understand the underlying biological mechanisms, this approach may be very helpful for the early diagnosis of various diseases such as cardiovascular disease in  clinical practice.
In order to achieve this long term goal, we need to deal with the following potential barriers in the current setup: 1) the data noise, and 2) the dataset size. Regarding the data noise, due to our straightforward and crude pulse acquisition equipment, the quality of the pulse waves is not satisfactory, so the measurement noise introduced during the collection needs to be quantified more carefully. As far as the dataset size, the current PPW dataset includes about 4,000 samples, which are not quite enough to train a high accuracy deep learning model.
 Our computational models will be further validated once more clinical data becomes available. {We will also explore the correlation between the PPW and time series data, such as the PPW versus the size of arterial plaque, in the future.}

\section{Acknowledgements}
This work is partially supported by the PSU-PKU Joint Center of
Computational Mathematics and Applications.  We would like thank Yuyan
Chen, Juncai He and Xiaodong Jia for many helpful discussions and
suggestions related to this paper and we would expecially like to
thank the following people for their generous efforts and helps in
collecting the PPW data:
Gang Chen,
Yujuan Chen,
Jing Feng,
Yanfeng, He,
Yaping He,
Bo Li,
Zichun Lian,
Shaobo Liang,
Jenny Li,
Jun Ma,
Dandan Pang,
Yanqiu Wang,
Bo Yang,
Duo Ye,
Xiaofeng Zhang,
and Shenglan Zhao.

\end{CJK*}

\end{document}